\documentclass[conference]{IEEEtran}
\IEEEoverridecommandlockouts

\usepackage{cite}
\usepackage{amsmath,amssymb,amsfonts}
\usepackage{algorithmic}
\usepackage{graphicx}
\usepackage{textcomp}
\usepackage{xcolor}
\usepackage{multirow}
\usepackage{booktabs}
\usepackage{array}
\usepackage{url}
\usepackage{pifont}
\usepackage{xspace}

\newcommand{\model}{CAMAR\xspace}

\DeclareRobustCommand*{\affiliationmark}[1]{\raisebox{0pt}[0pt][0pt]{\textsuperscript{\footnotesize\textit{#1}}}}

\newcolumntype{M}{c}
\newcommand{\hd}[1]{\makebox[34pt]{\scriptsize #1}}
\newcommand{\tightrules}{\setlength{\aboverulesep}{0.6pt}\setlength{\belowrulesep}{1pt}}

\def\BibTeX{{\rm B\kern-.05em{\sc i\kern-.025em b}\kern-.08em
    T\kern-.1667em\lower.7ex\hbox{E}\kern-.125emX}}

\begin{document}

\title{Multi-Modal Causal Disentanglement for\\Multi-Domain Sequential Audio Recommendation}

% Author block: IEEEtran_HOWTO "alternate long format" (recommended for more than three authors).
\author{\IEEEauthorblockN{Yu Ren\affiliationmark{a,b}\IEEEauthorrefmark{1},
Pengcheng Li\affiliationmark{b}\IEEEauthorrefmark{1},
Weijun Xu\affiliationmark{b},
Hengzhong Zhou\affiliationmark{b},
Shufeng Ye\affiliationmark{b},
Qiang Xiao\affiliationmark{b}\IEEEauthorrefmark{2},\\
Dongjing Wang\affiliationmark{a},
Chuanjiang Luo\affiliationmark{b},
Yong Li\affiliationmark{b}}
\IEEEauthorblockA{\affiliationmark{a}Hangzhou Dianzi University, Hangzhou, China\\
\affiliationmark{b}NetEase Cloud Music, Hangzhou, China\\
\{renyu05, dongjing.wang\}@hdu.edu.cn\\
\{lipengcheng06, xuweijun02, zhouhengzhong, yeshufeng, hzxiaoqiang, luochuanjiang03, liyong02\}@corp.netease.com}
\thanks{\IEEEauthorrefmark{1}These authors contributed equally to this work.}
\thanks{\IEEEauthorrefmark{2}Corresponding author.}}

\maketitle

\begin{abstract}
Modern music platforms expose users to heterogeneous audio content domains such as songs, playlists, and podcasts. Multi-domain sequential recommendation (MDSR) seeks to exploit such cross-domain behavior, yet two obstacles persist in practice. First, joint optimization across heterogeneous domains is prone to negative transfer, with high-traffic domains dominating the shared parameter space and suppressing weaker domains. Second, conventional ID-based collaborative signals fail to bridge domains that are semantically related but exhibit little behavioral overlap. This paper presents \model, a multi-modal causal disentanglement framework for multi-domain audio recommendation. From a causal perspective, user behavior is modeled as the joint outcome of a domain-invariant preference and domain-specific preferences. Guided by this view, \model factorizes user representations into domain-invariant and domain-specific components and enforces domain adversarial invariance together with orthogonality, restricting cross-domain interaction to the invariant subspace and suppressing harmful transfer paths. To strengthen cross-domain semantic coupling, \model derives trainable semantic-ID embeddings from pre-trained text, vision, and audio features through residual quantization, yielding an end-to-end multi-modal adaptation tailored to recommendation. A transfer network imputes user representations for inactive domains, mitigating cross-domain cold start. Experiments on both public and industrial datasets demonstrate consistent gains over strong baselines. In online A/B tests on NetEase Cloud Music, one of China's largest music streaming platforms, \model delivers consistent gains across all three domains in their main traffic scenarios: it increases play duration by $1.0\%$ on Song and $1.6\%$ on Podcast, and raises the like rate by $3.0\%$ on Playlist, validating its effectiveness in multi-domain audio recommendation at scale.
\end{abstract}


\section{Introduction}\label{sec:intro}

Modern music streaming platforms organize content into multiple co-existing domains---songs, curated playlists, and podcasts---among which users frequently switch within a single session~\cite{schedl2018current,van2013deep}. Such cross-domain consumption indicates that user preferences are simultaneously coupled and heterogeneous: the same user may exhibit very different interaction granularity, temporal patterns, and content focus across domains, while a relatively stable latent preference underlies all of them. Beyond behavior, this coupling is also reflected in content semantics, where topics, styles, and moods form a shared space across domains that cannot be captured by user--item interactions alone.

Conventional sequential recommenders~\cite{hidasi2015session,kang2018self,sun2019bert4rec} operate within a single domain and therefore cannot exploit cross-domain complementarity. Cross-domain sequential recommendation (CDSR) partially addresses this limitation by transferring sequential information from a source domain to a target domain~\cite{ma2019pi,cao2022contrastive,ma2024triple,lin2024mixed}. However, most CDSR methods adopt a target-centric, pairwise transfer paradigm and rely on item-ID collaborative signals for cross-domain alignment. When the recommendation scenario is extended to three or more domains, the number of pairwise relations grows combinatorially, and pairwise alignment yields fragmented knowledge sharing without a unified fusion mechanism.

Multi-domain sequential recommendation (MDSR)~\cite{park2023cracking,hwang2024multi} mitigates these issues by jointly modeling all domains under a unified framework. Each domain serves both as a prediction target and as an information source for the others. This symmetric formulation introduces richer cross-domain signals but also amplifies the risk of negative transfer~\cite{zhang2022survey}, since harmful information now propagates globally through shared structures rather than locally between two domains.

Despite recent progress, two challenges remain open in MDSR.

\noindent\textbf{Challenge 1: Negative transfer and domain dominance under multi-domain coupling.} In MDSR, all domains share parts of the representation space and parameters, turning the local cross-domain transfer of CDSR into a global coupling~\cite{ju2025revisiting,park2024pacer,park2023cracking}. When domains differ substantially in interaction density, behavior patterns, or semantic distributions, irrelevant or harmful signals are repeatedly amplified across domains~\cite{park2023cracking,park2024pacer}. This effect is particularly pronounced in music platforms: the song domain typically attracts far more interactions than playlists or podcasts. The gradient of shared parameters is therefore dominated by the high-traffic domain, biasing the shared representation space and systematically degrading recommendation quality on the lower-traffic domains.

\noindent\textbf{Challenge 2: Insufficient cross-domain content semantics.} Existing MDSR methods rely primarily on item-ID collaborative signals, whose effectiveness is conditioned on the existence of users active in multiple domains. In real-world platforms, however, such multi-domain active users constitute only a small fraction of the user base. Without sufficient ID co-occurrence, traditional methods cannot establish reliable cross-domain item associations. Multi-modal content semantics (text, vision, and audio) provide a more general bridge: items from different domains can still be aligned in the semantic space even when they are never consumed by the same user~\cite{hou2022towards,rajput2023recommender,yang2024cascading}. Effectively integrating such content semantics for cross-domain modeling has received limited attention.

\noindent\textbf{Proposed framework.} This paper introduces \model, a multi-modal multi-domain sequential recommendation framework for music platforms. To address \textbf{Challenge~1}, \model adopts a causal view of multi-domain preference generation: user behavior is jointly driven by a domain-invariant preference (e.g.,\ topical or affective preferences such as ``healing'' or ``commuting companion'') and a set of domain-specific preferences (e.g.,\ timbre and rhythm in the song domain, narrative pacing in the podcast domain). \model materializes this generative structure through a structural causal graph that identifies three spurious paths responsible for negative transfer. To cut these paths, \model decomposes user representations into domain-invariant and domain-specific factors under three constraints: orthogonality between the two factors, domain adversarial training on the invariant factor, and architectural restriction of cross-domain interaction. To address \textbf{Challenge~2}, \model converts pre-trained multi-modal features into compact semantic IDs using an offline residual-quantized variational autoencoder (RQ-VAE) and maintains a trainable, cross-domain shared codebook within the recommender. This design aligns items from different domains in the semantic space without requiring behavioral co-occurrence. Finally, a lightweight cross-domain transfer network (TransNet) imputes user representations on domains where the user is inactive, alleviating the cold-start and partial-observation issues common in multi-domain settings.

\noindent\textbf{Contributions.} The contributions of this paper are summarized as follows.
\begin{itemize}
    \item We propose \model, a unified multi-modal multi-domain sequential recommendation framework, and introduce a causal formulation of multi-domain preference generation that decomposes user preference into domain-invariant and domain-specific factors. Guided by the resulting causal graph, we derive three complementary constraints to jointly suppress negative transfer in multi-domain joint optimization.
    \item We design a trainable cross-domain semantic-ID codebook that integrates text, vision, and audio semantics into the recommender, aligning items across heterogeneous audio domains without behavioral co-occurrence, and further introduce a lightweight transfer network that imputes user representations for inactive domains to alleviate cross-domain cold start.
    \item Extensive offline experiments on two real-world datasets and online A/B tests verify the effectiveness of the proposed method.
\end{itemize}

\section{Related Work}\label{sec:related}

\subsection{Sequential Recommendation}
Sequential recommendation predicts the next user action conditioned on historical interactions. Early approaches build on Markov chains or recurrent networks such as GRU4Rec~\cite{hidasi2015session}, which capture short- and long-term interest with gated units but suffer from limited efficiency and degraded long-range dependency in long sequences. Self-attention models---SASRec~\cite{kang2018self} and BERT4Rec~\cite{sun2019bert4rec}, both built on the Transformer architecture~\cite{vaswani2017attention}---have since become the dominant paradigm. Beyond single-paradigm encoders, session- and graph-based formulations inject high-order collaborative structure into sequential modeling: session-based models exploit local invariance~\cite{chen2019session} and fine-grained interaction graphs~\cite{han2023fm} to capture intra-session item transitions, while graph collaborative optimization~\cite{xiao2023graph} couples sequential behavior with graph collaborative signals, and a cascaded cross-attention model~\cite{huang2023cascaded} further fuses review semantics with behavior sequences. More recently, Mamba4Rec~\cite{liu2024mamba4rec} replaces self-attention with linear-time state-space modeling for more efficient sequence encoding. These efforts focus mainly on single-domain settings and leave the joint modeling of user preferences across multiple domains largely open.

\subsection{Cross-Domain and Multi-Domain Sequential Recommendation}
Cross-domain recommendation (CDR) leverages auxiliary signals from related domains to mitigate sparsity and cold start in a target domain. Classical CDR methods such as CMF~\cite{singh2008relational}, EMCDR~\cite{man2017cross}, and DTCDR~\cite{zhu2019dtcdr} transfer knowledge through shared latent factors or feature mappings. More recently, some CDR methods exploit multi-modal data to further boost transfer, e.g.,\ MuSiC~\cite{zhang2025leveraging} combines multi-modal large language models with cross-domain diffusion generation, and P2M2-CDR~\cite{wang2024privacy} disentangles multi-modal features under a privacy-preserving transfer scheme. All the above methods, however, treat user preferences as static. CDSR extends cross-domain transfer with temporal dynamics by modeling users' chronologically ordered interactions. Existing CDSR methods typically rely on gated transfer modules~\cite{ma2019pi}, graph neural networks (e.g.,\ C$^2$DSR~\cite{cao2022contrastive}, BiTGCF~\cite{liu2020cross}, MIFN~\cite{ma2022mixed}), mixed-attention encoders~\cite{lin2024mixed}, or Transformer-based cross-domain encoders~\cite{ma2024triple}. Most of these models are inherently pairwise and scale poorly when the number of domains exceeds two. To address this scalability barrier, MDSR jointly models multiple domains within a unified sequential framework. CGRec~\cite{park2023cracking} formulates cross-domain knowledge sharing as a cooperative game and reweights domains via Shapley values; MDSR-DSL~\cite{hwang2024multi} learns per-domain spaces and a domain-to-domain adaptation module; SyNCRec~\cite{park2024pacer} reconciles single-domain and cross-domain branches to preserve domain-specific patterns; AutoCDSR~\cite{ju2025revisiting} introduces a Pareto-optimal self-attention to suppress harmful information flow; and MOSAIC~\cite{bahi2026mosaic} factorizes user preferences into orthogonal components through a triple-encoder. PMMRec~\cite{li2024multi} further discards item IDs altogether and learns transferable multi-modal representations for cross-platform recommendation. These methods focus mostly on structural information sharing and largely overlook the explicit causal structure underlying multi-domain preference generation.

\subsection{Causal Inference and Disentangled Representations in Recommendation}
Causal inference has been adopted in recommendation primarily for bias correction (e.g.,\ inverse propensity scoring~\cite{schnabel2016recommendations}) and for invariant learning~\cite{arjovsky2019invariant}. In cross-domain recommendation, DisenCDR~\cite{cao2022disencdr} learns disentangled shared and domain-specific user representations from a mutual-information perspective. Grace~\cite{zhang2023connecting} casts cross-domain modeling as causal invariant learning, treating user behavior as the joint outcome of invariant and domain-specific factors. C$^2$DR~\cite{menglin2024c2dr} constructs a causal graph for cross-domain recommendation and derives independence constraints from it to address negative transfer, while CausalInt~\cite{wang2022causalint} cuts harmful causal paths in multi-scenario modeling through interventions. These methods validate the value of causal disentanglement but largely focus on the two-domain setting; complex cross-domain dependencies in MDSR remain under-modeled.

\section{Preliminaries}\label{sec:prelim}

This paper considers multi-domain sequential recommendation in which a user's history may span an arbitrary number of domains. Let $\mathcal{D}=\{d_{1},d_{2},\dots\}$ denote the set of domains with $|\mathcal{D}|\ge 3$. For domain $d\in\mathcal{D}$, the per-domain interaction sequence of user $u$ is $X_u^{d}=[x_{1}^{d},x_{2}^{d},\dots]$, where $x_{m}^{d}\in\mathcal{V}^{d}$ and $\mathcal{V}^{d}$ is the item set of domain $d$. The multi-domain sequence $X_u=[x_{1},x_{2},\dots,x_{M}]$ is obtained by ordering the items from all domains chronologically, with $x_{m}\in\mathcal{V}=\bigcup_{d\in\mathcal{D}}\mathcal{V}^{d}$ and $M=\sum_{d\in\mathcal{D}}|X_u^{d}|$. We focus on the recall task: given both the chronological multi-domain history $X_u$ and the per-domain subsequences $\{X_u^{d}\}_{d\in\mathcal{D}}$, the model identifies the next item $x^{\star}$ that user $u$ will interact with in each target domain $d\in\mathcal{D}$,
\begin{equation}
\arg\max_{x^{\star}\in\mathcal{V}^{d}}\ P\!\left(x^{\star}\,\bigm|\,\{X_u^{d}\}_{d\in\mathcal{D}}\right),
\label{eq:obj}
\end{equation}
which must hold simultaneously for every $d\in\mathcal{D}$.

\section{Methodology}\label{sec:method}

\begin{figure*}[t]
\centering
\includegraphics[width=\textwidth]{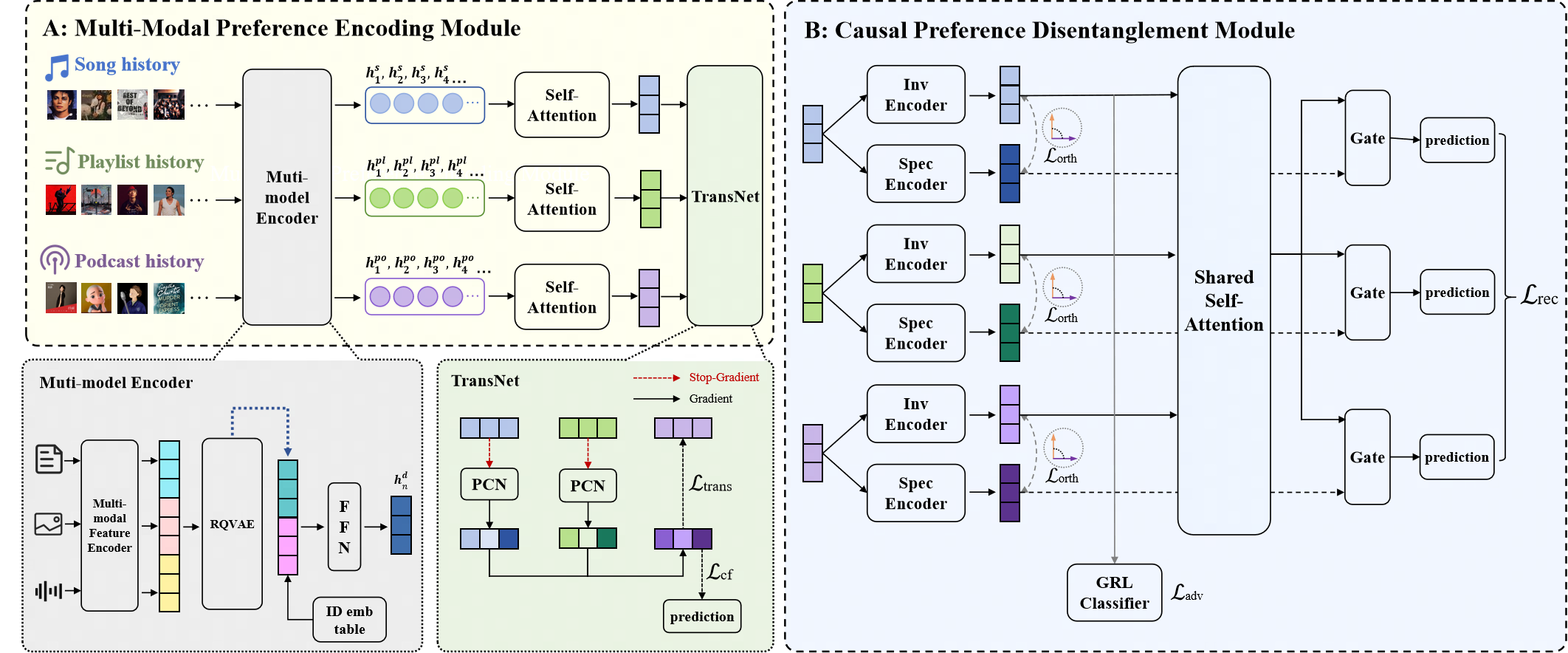}
\caption{The overall workflow of \model, an end-to-end model consisting of two sequentially connected stages: (A) the Multi-Modal Preference Encoding Module encodes each domain's behavior sequence into a complete per-domain user representation, and (B) the Causal Preference Disentanglement Module disentangles each representation into invariant and domain-specific preferences and fuses them for scoring.}
\label{fig:framework}
\end{figure*}

This section presents \model. As illustrated in Fig.~\ref{fig:framework}, given the per-domain interaction sequences of a user, \model produces per-domain recall scores through two main modules. Section~\ref{sec:enc} describes the multi-modal preference encoding module, which constructs domain-level user representations: starting from raw multi-modal item features, \model performs item fusion, intra-domain sequential encoding, and cross-domain imputation, yielding a complete set $\{\tilde{\mathbf{h}}_{u}^{d}\}_{d\in\mathcal{D}}$ in which every domain admits a representation. Section~\ref{sec:causal} describes the causal preference disentanglement module, which performs invariant--specific decomposition and controlled cross-domain fusion on top of $\{\tilde{\mathbf{h}}_{u}^{d}\}$, producing the final per-domain user vector $\mathbf{r}_{u}^{d}$ for scoring. Section~\ref{sec:train} specifies the training and inference procedures.

\subsection{Multi-Modal Preference Encoding Module}\label{sec:enc}

\subsubsection{Cross-Domain Multi-Modal Semantic IDs}\label{sec:semid}
For every item $i\in\mathcal{V}$, pre-trained encoders extract content features from three modalities: a text vector $\mathbf{x}_{i}^{t}$ produced by Qwen3-Embedding-0.6B~\cite{zhang2025qwen3} over the title, description, and tags; a visual vector $\mathbf{x}_{i}^{v}$ produced by a Vision Transformer~\cite{dosovitskiy2020image} over the cover or poster image; and an audio vector $\mathbf{x}_{i}^{a}$ produced by a publicly available audio representation model~\cite{li2024mert} over a song clip or a podcast snippet. The unified content vector $\mathbf{x}_{i}^{\mathrm{mm}}=[\mathbf{x}_{i}^{t};\mathbf{x}_{i}^{v};\mathbf{x}_{i}^{a}]$ is obtained by feature-wise concatenation. Following the semantic-ID line of work~\cite{rajput2023recommender}, a single RQ-VAE~\cite{lee2022autoregressive} is then jointly trained on the union of items from all domains $\mathcal{V}$ to discretize $\mathbf{x}_{i}^{\mathrm{mm}}$ into a length-$L$ semantic ID $S_i=(c_i^{(1)},\ldots,c_i^{(L)})$ with $c_i^{(l)}\in\{1,\ldots,K_c\}$, where $c_i^{(l)}$ is the codeword index assigned to item $i$ at the $l$-th quantization level and $K_c$ is the per-level codebook size. A shared RQ-VAE---rather than per-domain or per-modality codebooks---encourages items with similar topics, moods, or styles across domains to share the leading codewords, providing a stable semantic anchor for the downstream causal disentanglement. The quantization is performed offline; the recommender consumes only the discrete $S_i$.

\subsubsection{Trainable Semantic Codebook and Item Representation}\label{sec:item}
For each quantization level $l\in\{1,\dots,L\}$, \model maintains a trainable codebook embedding table $\mathbf{E}^{(l)}\in\mathbb{R}^{K_c\times D_c}$, where $D_c$ is the semantic embedding dimension. The codebook is shared across all domains. The multi-modal semantic embedding of item $i$ is obtained by summing the looked-up codeword embeddings across levels:
\begin{equation}
\mathbf{e}_{\mathrm{sem},i}=\sum_{l=1}^{L}\mathrm{onehot}\!\bigl(c_i^{(l)}\bigr)\,\mathbf{E}^{(l)},\qquad \mathbf{e}_{\mathrm{sem},i}\in\mathbb{R}^{D_c},
\label{eq:sem}
\end{equation}
where $\mathrm{onehot}(c)\in\{0,1\}^{K_c}$ is the one-hot vector with a single $1$ at position $c$. Sharing the codebook ensures that items from different domains receive comparable semantic coordinates at the input layer.

To preserve domain-specific collaborative patterns, every item $i$ in domain $d$ additionally carries a domain-specific ID embedding $\mathbf{e}_{\mathrm{id},i}^{d}\in\mathbb{R}^{D_{\mathrm{id}}}$. The two streams are fused through a domain-specific feed-forward network:
\begin{equation}
\mathbf{e}_{i}^{d}=\mathrm{FFN}^{d}\!\bigl([\mathbf{e}_{\mathrm{id},i}^{d};\mathbf{e}_{\mathrm{sem},i}]\bigr)\in\mathbb{R}^{D},
\label{eq:item}
\end{equation}
where $D$ is the embedding dimension. The cross-domain shared component is concentrated in $\mathbf{e}_{\mathrm{sem},i}$, while $\mathbf{e}_{\mathrm{id},i}^{d}$ and $\mathrm{FFN}^{d}$ carry the domain-specific signal.

\subsubsection{Intra-Domain Sequential Encoding}\label{sec:seq}
For user $u$ and domain $d$, the fused embedding $\mathbf{e}_{i}^{d}$ of every item $i\in X_u^{d}$ is added to a learnable position embedding and stacked into the sequence matrix $\mathbf{H}_{u}^{d}\in\mathbb{R}^{n\times D}$ ($n$ denoting the truncated sequence length). $\mathbf{H}_{u}^{d}$ is fed into a stack of multi-head self-attention (MHA) layers and the per-position outputs are aggregated by attention pooling:
\begin{equation}
\mathbf{h}_{u}^{d}=\mathrm{AttnPool}\!\left(\mathrm{MHA}^{d}(\mathbf{H}_{u}^{d})\right),
\label{eq:hud}
\end{equation}
where $\mathrm{MHA}^{d}(\cdot)$ is a multi-head self-attention stack and $\mathrm{AttnPool}(\cdot)$ aggregates per-position outputs into a single vector via learned attention weights. Each domain owns an independent encoder and position embedding.

\subsubsection{Cross-Domain Representation Imputation}\label{sec:transnet}
Let $\mathcal{D}_{u}^{+}\subseteq\mathcal{D}$ denote the set of domains in which user $u$ is active. Equation~\eqref{eq:hud} only produces $\mathbf{h}_{u}^{d}$ for $d\in\mathcal{D}_{u}^{+}$. In real-world music streaming platforms, a large fraction of users interact with only one or two domains, leaving the remaining domains without any behavioral signal. For these inactive domains, the model cannot generate any recommendation. TransNet addresses this cold-start issue by imputing the missing domain representations from the user's active domains, enabling \model to produce recall results even for domains where the user has no historical interaction.

TransNet consists of a set of Preference Completion Networks (PCNs), one per ordered domain pair. For every ordered pair $(d_s,d_t)$ with $d_s\neq d_t$, where $d_s$ is the source domain and $d_t$ is the target domain,
\begin{equation}
\hat{\mathbf{h}}_{u}^{d_s\rightarrow d_t}=\mathrm{PCN}_{d_s\rightarrow d_t}\!\bigl(\mathrm{sg}(\mathbf{h}_{u}^{d_s})\bigr),
\label{eq:trans}
\end{equation}
where $\mathrm{sg}(\cdot)$ is the stop-gradient operator and each $\mathrm{PCN}_{d_s\rightarrow d_t}$ is a two-layer ReLU MLP with parameters dedicated to the directed pair. The stop-gradient serves two purposes: it prevents imputation losses from distorting the already-learned source-domain encoder $\mathrm{MHA}^{d_s}$, and it decouples the optimization of PCN from the main recommendation objective, avoiding gradient interference between imputation and recall. The completed domain representation is
\begin{equation}
\tilde{\mathbf{h}}_{u}^{d}=
\begin{cases}
\mathbf{h}_{u}^{d}, & d\in\mathcal{D}_{u}^{+},\\[2pt]
\dfrac{1}{|\mathcal{D}_{u}^{+}|}\displaystyle\sum_{d_s\in\mathcal{D}_{u}^{+}}\hat{\mathbf{h}}_{u}^{d_s\rightarrow d}, & d\notin\mathcal{D}_{u}^{+}.
\end{cases}
\label{eq:tildeh}
\end{equation}
After this step, every domain admits a representation $\tilde{\mathbf{h}}_{u}^{d}\in\mathbb{R}^{D}$ of the same dimension. The pairwise design instantiates $|\mathcal{D}|(|\mathcal{D}|-1)$ PCNs, which is affordable here since each PCN is a two-layer MLP that is negligible in size relative to the item embedding table. For platforms with many domains, the pairwise PCNs can be replaced by one shared network modulated by gates generated from learnable domain embeddings~\cite{chang2023pepnet}, reducing the parameter growth to $O(|\mathcal{D}|)$; we retain the pairwise form under three domains as it is easier to optimize.

\noindent\textbf{TransNet supervision.} Let $\mathcal{P}$ be the set of ordered domain pairs that contain overlapping users and let $\mathcal{U}_{d_s\cap d_t}=\mathcal{U}_{d_s}\cap\mathcal{U}_{d_t}$ denote the user set active in both $d_s$ and $d_t$, where $\mathcal{U}_d$ is the set of users active in domain $d$. \model supervises TransNet at two levels. The representation-level matching loss aligns the imputed representation with the true target-domain representation,
\begin{equation}
\mathcal{L}_{\mathrm{rep}}=\frac{1}{|\mathcal{P}|}\sum_{(d_s,d_t)\in\mathcal{P}}\mathbb{E}_{u\in\mathcal{U}_{d_s\cap d_t}}\!\left[\bigl\|\hat{\mathbf{h}}_{u}^{d_s\rightarrow d_t}-\mathrm{sg}(\mathbf{h}_{u}^{d_t})\bigr\|_2^2\right].
\label{eq:Lrep}
\end{equation}
Stop-gradient on the target side ensures that $\mathcal{L}_{\mathrm{rep}}$ only updates TransNet. The task-level cold-start loss enforces end-to-end consistency on the recall objective. For an overlapping user $u\in\mathcal{U}_{d_s\cap d_t}$, the target-domain history is masked during training and $\hat{\mathbf{h}}_{u}^{d_s\rightarrow d_t}$ replaces $\mathbf{h}_{u}^{d_t}$ in the downstream pipeline, yielding a cold-start user vector $\mathbf{r}_{\mathrm{cf},u}^{d_t}$ that is supervised by the user's ground-truth next interaction $x_{u}^{d_t,+}$ in domain $d_t$,
\begin{equation}
\mathcal{L}_{\mathrm{cf}}=\frac{1}{|\mathcal{P}|}\sum_{(d_s,d_t)\in\mathcal{P}}\mathbb{E}_{u\in\mathcal{U}_{d_s\cap d_t}}\!\left[\ell_{\mathrm{rec}}^{d_t}\!\bigl(\mathbf{r}_{\mathrm{cf},u}^{d_t},\,x_{u}^{d_t,+}\bigr)\right],
\label{eq:Lcf}
\end{equation}
where $\ell_{\mathrm{rec}}^{d_t}(\cdot,\cdot)$ is the per-user sampled-softmax recall loss in domain $d_t$ (cf.\ Eq.~\eqref{eq:Lrec}). Because the source input is stopped in Eq.~\eqref{eq:trans}, $\mathcal{L}_{\mathrm{cf}}$ does not back-propagate into the source encoder. The two terms are combined into the TransNet objective $\mathcal{L}_{\mathrm{TN}}=\mathcal{L}_{\mathrm{rep}}+\mu\,\mathcal{L}_{\mathrm{cf}}$, where $\mu\in[0,1]$ controls the relative weight of task-level supervision.

\subsection{Causal Preference Disentanglement Module}\label{sec:causal}

\subsubsection{Structural Causal Graph for Multi-Domain Preference}\label{sec:scm}
\begin{figure}[t]
\centerline{\includegraphics[width=0.9\columnwidth]{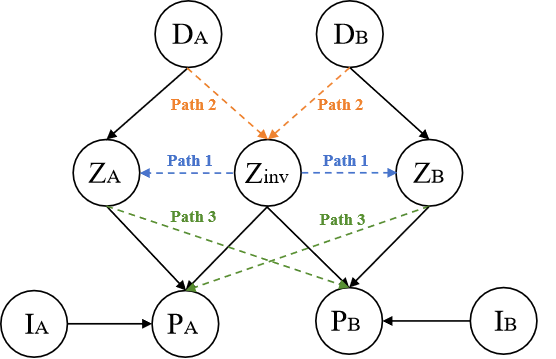}}
\caption{Structural causal graph of multi-domain preference generation. Solid arrows denote desired generative paths; dashed arrows denote spurious or confounding paths (Paths~1--3) to be cut.}
\label{fig:scm}
\end{figure}

The structural causal model is illustrated with two domains $A$ and $B$ in Fig.~\ref{fig:scm}; the analysis generalizes to arbitrary $|\mathcal{D}|$. Five groups of variables are involved: domain attributes $D_A,D_B$ encoding content modality, interaction granularity, and behavior distribution; candidate items $I_A,I_B$; domain-specific preferences $Z_A,Z_B$; the cross-domain domain-invariant preference $Z_{\mathrm{inv}}$; and the recall scores $P_A,P_B$.

Solid arrows denote desired generative paths. Domain attributes shape domain-specific preferences ($D_A\!\rightarrow\!Z_A$, $D_B\!\rightarrow\!Z_B$): the song domain emphasizes timbre and rhythm, the playlist domain emphasizes music genres and listening contexts, whereas the podcast domain emphasizes topic and narrative, reflecting the distinct preference patterns across domains. Domain-specific preferences drive within-domain prediction ($Z_A\!\rightarrow\!P_A$, $Z_B\!\rightarrow\!P_B$). The domain-invariant preference $Z_{\mathrm{inv}}$ simultaneously affects all domains ($Z_{\mathrm{inv}}\!\rightarrow\!P_A,P_B$) and encodes high-level affective or topical preferences (e.g.,\ ``healing'', ``commute'') that manifest consistently across domains. Candidate items influence only their own domain ($I_A\!\rightarrow\!P_A$, $I_B\!\rightarrow\!P_B$).

Dashed arrows denote spurious or confounding paths to be cut, each labeled in Fig.~\ref{fig:scm}:
\begin{itemize}
\item \textbf{Path 1}: $Z_{\mathrm{inv}}\!\rightarrow\!Z_A,\,Z_B$.\ \ Invariant information contaminates domain-specific representations, causing $Z_A$ and $Z_B$ to lose their role as domain residuals.
\item \textbf{Path 2}: $D_A,D_B\!\rightarrow\!Z_{\mathrm{inv}}$.\ \ Domain attributes leak into the domain-invariant preference, biasing $Z_{\mathrm{inv}}$ towards high-traffic domains.
\item \textbf{Path 3}: $Z_A\!\rightarrow\!P_B,\,Z_B\!\rightarrow\!P_A$.\ \ Domain-specific signals interfere with other domains' predictions.
\end{itemize}
Sections~\ref{sec:decomp}--\ref{sec:fuse} present the mechanisms that cut Paths~1--3, respectively.

\subsubsection{Invariant--Specific Decomposition}\label{sec:decomp}
This subsection addresses \textbf{Path~1} ($Z_{\mathrm{inv}}\!\rightarrow\!Z_A,Z_B$) by decomposing each domain representation into an invariant and a specific component and enforcing their orthogonality.

Two domain-specific single-layer linear projections extract the invariant and specific components from $\tilde{\mathbf{h}}_{u}^{d}$:
\begin{equation}
\mathbf{z}_{\mathrm{inv},u}^{d}=\mathrm{InvEnc}^{d}\!\bigl(\tilde{\mathbf{h}}_{u}^{d}\bigr),\quad
\mathbf{z}_{\mathrm{spec},u}^{d}=\mathrm{SpecEnc}^{d}\!\bigl(\tilde{\mathbf{h}}_{u}^{d}\bigr),
\label{eq:decomp}
\end{equation}
where $\tilde{\mathbf{h}}_{u}^{d}$ is the completed domain representation from Eq.~\eqref{eq:tildeh}, and $\mathbf{z}_{\mathrm{inv},u}^{d},\mathbf{z}_{\mathrm{spec},u}^{d}\in\mathbb{R}^{D}$. The two branches are distinguished by the training objectives rather than by architecture: the domain adversarial loss is applied to $\mathbf{z}_{\mathrm{inv},u}^{d}$ alone, driving it towards a domain-indistinguishable subspace that is further aligned across domains by the constraints in Section~\ref{sec:dedom}, whereas $\mathbf{z}_{\mathrm{spec},u}^{d}$ is excluded from any cross-domain interaction and is thus left to capture the complementary domain-internal residual. To enforce information-level separation between the two components within each domain, an orthogonality regularizer is applied over all users $\mathcal{U}_{d}$ active in each domain:
\begin{equation}
\mathcal{L}_{\mathrm{ortho}}=\frac{1}{|\mathcal{D}|}\sum_{d\in\mathcal{D}}\mathbb{E}_{u\in\mathcal{U}_{d}}\!\left[\bigl\langle\mathbf{z}_{\mathrm{inv},u}^{d},\,\mathbf{z}_{\mathrm{spec},u}^{d}\bigr\rangle^{2}\right].
\label{eq:Lortho}
\end{equation}
Under this constraint, $\mathbf{z}_{\mathrm{inv}}$ encodes the domain-invariant preference (cross-domain reusable factors such as topic, mood, and style), while $\mathbf{z}_{\mathrm{spec}}$ encodes the domain-specific preference (domain-internal residuals such as timbre composition or narrative pacing).

\subsubsection{De-Domainization via Domain Adversarial Training}\label{sec:dedom}
This subsection addresses \textbf{Path~2} ($D_A,D_B\!\rightarrow\!Z_{\mathrm{inv}}$) by removing domain-identifiable information from the invariant representation through adversarial training.

A gradient reversal layer (GRL)~\cite{ganin2016domain} $\mathrm{GRL}_{\lambda}$ is inserted before $\mathbf{z}_{\mathrm{inv}}$ enters the shared self-attention. $\mathrm{GRL}_{\lambda}$ acts as the identity in the forward pass and scales the gradient by $-\lambda$ in the backward pass. The output is fed into a domain classifier $\mathrm{Classifier}(\cdot)$, a two-layer MLP shared across domains, which produces a $|\mathcal{D}|$-way distribution:
\begin{equation}
\hat{\mathbf{p}}_{u}^{d}=\mathrm{Classifier}\!\bigl(\mathrm{GRL}_{\lambda}(\mathbf{z}_{\mathrm{inv},u}^{d})\bigr).
\label{eq:grl}
\end{equation}
The coefficient $\lambda$ is linearly warmed up from $0$ to $1$ during training. The domain adversarial loss is the standard multi-class cross-entropy
\begin{equation}
\mathcal{L}_{\mathrm{adv}}=-\frac{1}{N|\mathcal{D}|}\sum_{u=1}^{N}\sum_{d\in\mathcal{D}}\log [\hat{\mathbf{p}}_{u}^{d}]_{d},
\label{eq:Ladv}
\end{equation}
where $[\hat{\mathbf{p}}_{u}^{d}]_{d}$ denotes the probability that the classifier assigns to the true domain label $d$ (i.e.,\ the $d$-th component of $\hat{\mathbf{p}}_{u}^{d}$ from Eq.~\eqref{eq:grl}) and $N$ is the batch size. Through GRL, the classifier and the invariant encoder play a min-max game: the classifier seeks to identify the source domain of $\mathbf{z}_{\mathrm{inv}}$, while the encoder is pushed towards a representation under which the source domain is indistinguishable. At equilibrium, $P(\mathbf{z}_{\mathrm{inv}}\mid D=d_i)\approx P(\mathbf{z}_{\mathrm{inv}}\mid D=d_j)$.

\subsubsection{Structured Cross-Domain Interaction and Prediction}\label{sec:fuse}
This subsection addresses \textbf{Path~3} ($Z_A\!\rightarrow\!P_B,\,Z_B\!\rightarrow\!P_A$) by restricting cross-domain interaction to the invariant subspace only.

Only invariant representations enter the cross-domain interaction. The invariant components are stacked into a token sequence $\mathbf{Z}_{\mathrm{inv},u}=[\mathbf{z}_{\mathrm{inv},u}^{d_1},\ldots,\mathbf{z}_{\mathrm{inv},u}^{d_{|\mathcal{D}|}}]$ and processed by a shared self-attention layer with an active-domain mask:
\begin{align}
[\bar{\mathbf{z}}_{\mathrm{inv},u}^{d_1},\ldots,\bar{\mathbf{z}}_{\mathrm{inv},u}^{d_{|\mathcal{D}|}}]&=\mathrm{SharedAttn}\!\bigl(\mathbf{Z}_{\mathrm{inv},u};\mathbf{M}_{u}^{\mathrm{active}}\bigr),\nonumber\\
\mathbf{z}_{\mathrm{share},u}&=\frac{1}{|\mathcal{D}_{u}^{+}|}\sum_{d\in\mathcal{D}_{u}^{+}}\bar{\mathbf{z}}_{\mathrm{inv},u}^{d},
\label{eq:share}
\end{align}
where $\bar{\mathbf{z}}_{\mathrm{inv},u}^{d}$ denotes the invariant representation of domain $d$ after cross-domain attention, and $\mathbf{z}_{\mathrm{share},u}$ is the aggregated cross-domain shared preference. The mask $\mathbf{M}_{u}^{\mathrm{active}}$ enables bidirectional attention only among active domains in $\mathcal{D}_{u}^{+}$. For an inactive domain whose $\mathbf{z}_{\mathrm{inv},u}^{d}$ is imputed by TransNet, attention is bypassed by setting $\bar{\mathbf{z}}_{\mathrm{inv},u}^{d}=\mathbf{z}_{\mathrm{inv},u}^{d}$, and the average pooling is taken only over active domains, ensuring that imputed signals do not contaminate the shared preference. By construction, $\mathbf{z}_{\mathrm{spec},u}^{d}$ never leaves its own domain branch, which structurally guarantees $Z_{d_i}\!\not\rightarrow\!P_{d_j}$ for $d_i\neq d_j$.

A gating mechanism adaptively fuses the cross-domain shared preference with the per-domain specific preference,
\begin{align}
\mathbf{r}_{u}^{d}&=\mathcal{N}_{2}\!\Bigl(\mathbf{g}_{u}^{d}\odot\mathbf{z}_{\mathrm{share},u}+(1-\mathbf{g}_{u}^{d})\odot\mathbf{z}_{\mathrm{spec},u}^{d}\Bigr),\nonumber\\
\mathbf{g}_{u}^{d}&=\sigma\!\bigl(\mathbf{W}_{g}^{d}[\mathbf{z}_{\mathrm{share},u};\mathbf{z}_{\mathrm{spec},u}^{d}]+\mathbf{b}_{g}^{d}\bigr),
\label{eq:gate}
\end{align}
where $\mathbf{g}_{u}^{d}\in\mathbb{R}^{D}$ is the gate, $\mathbf{W}_{g}^{d}\in\mathbb{R}^{D\times 2D}$ and $\mathbf{b}_{g}^{d}\in\mathbb{R}^{D}$ are learnable parameters, $\sigma(\cdot)$ is the element-wise sigmoid, $\odot$ is element-wise multiplication, and $\mathcal{N}_{2}(\cdot)$ denotes $\ell_2$ normalization. The recall score for domain $d$ is defined as $s(i)=(\mathbf{r}_{u}^{d})^{\top}\mathbf{e}_{i}^{d}$, where $u$ and $d$ are determined by context. The multi-domain recall loss is
\begin{equation}
\mathcal{L}_{\mathrm{rec}}=-\sum_{d\in\mathcal{D}}\frac{1}{|\mathcal{U}_{d}|}\sum_{u\in\mathcal{U}_{d}}\log\frac{\exp s(i^{+})}{\exp s(i^{+})+\sum\limits_{i^{-}\in\mathcal{S}_{u}^{-}}\exp s(i^{-})},
\label{eq:Lrec}
\end{equation}
where $i^{+}$ is the ground-truth next item and $\mathcal{S}_{u}^{-}$ is the per-user sampled negative set.

\subsection{Training and Inference}\label{sec:train}

\noindent\textbf{Training objective.} The full training objective combines the recall loss, the TransNet supervision, and the two causal-disentanglement losses,
\begin{equation}
\mathcal{L}=\mathcal{L}_{\mathrm{rec}}+\alpha\,\mathcal{L}_{\mathrm{TN}}+\beta\,\mathcal{L}_{\mathrm{ortho}}+\gamma\,\mathcal{L}_{\mathrm{adv}},
\label{eq:total}
\end{equation}
where $\alpha,\beta,\gamma\in[0,1]$. Training proceeds in three stages for stability: $\mathcal{L}_{\mathrm{rec}}$ is optimized alone in the first stage to stabilize the recall space; $\mathcal{L}_{\mathrm{ortho}}$ is then introduced; finally $\mathcal{L}_{\mathrm{adv}}$ and the task-level TransNet supervision $\mathcal{L}_{\mathrm{TN}}$ are activated.

\noindent\textbf{Inference.} The domain classifier, GRL, and the disentanglement loss branches ($\mathcal{L}_{\mathrm{ortho}}$, $\mathcal{L}_{\mathrm{adv}}$) are removed at inference. TransNet is retained to impute domain representations for cold-start users who lack history in certain domains. Given the user history, \model executes the forward path of Sections~\ref{sec:enc}--\ref{sec:causal} to obtain $\mathbf{r}_{u}^{d}$ for every target domain. The top-$K$ items in $\mathcal{V}^{d}$ ranked by $s(i)$ are returned as the recall result for domain $d$.

\section{Experiments}\label{sec:exp}

In this section, we conduct a variety of offline experiments and online tests to evaluate the proposed method. Specifically, our aim is to address the following research questions:
\begin{itemize}
\item \textbf{RQ1.} How does \model perform against state-of-the-art single-domain and multi-domain baselines?
\item \textbf{RQ2.} What is the contribution of each component (causal disentanglement, TransNet, multi-modal semantic IDs)?
\item \textbf{RQ3.} How well does \model transfer knowledge to serve users who are cold in a target domain?
\item \textbf{RQ4.} Does \model deliver consistent gains in an industrial online A/B test?
\end{itemize}

\subsection{Experimental Setup}

\subsubsection{Datasets}
Two datasets covering three domains each are used.

\noindent\textbf{Industrial Dataset.} The Industrial Dataset is collected from anonymized production logs of NetEase Cloud Music, one of the largest music streaming platforms in China, and covers users' full-lifecycle interaction history across three content domains: Song, Playlist, and Podcast. Specifically, interactions in the Song domain are constructed from users' like behaviors, interactions in the Playlist domain are constructed from like, completion, and favorite behaviors, and interactions in the Podcast domain are constructed from playback and favorite behaviors. All records are anonymized at the user level, and only aggregated statistics are reported. To obtain a clean multi-domain benchmark, only users who are active in all three domains are retained, yielding a shared user set of approximately $3.5\!\times\!10^{6}$ users. For each retained user, all observed (user, item) interactions form the cross-domain interaction history. For each domain, the earliest interaction occurring after the history cutoff point is used as the recall ground truth.

\noindent\textbf{Amazon Reviews (public).} To evaluate the generalizability of the proposed framework beyond the audio domain, we additionally adopt the Amazon Reviews dataset as a public benchmark. It contains three categories: Books, Kindle\_Store, and Movies\&TV. In this dataset, we use only the textual information of items as item features. Users active in at least two domains with no fewer than five interactions per active domain are retained. For each retained user, the last interaction in each active domain is taken as the positive sample, while the preceding interactions constitute the cross-domain input history.

Statistics of both datasets are summarized in Table~\ref{tab:stats}.

\begin{table}[t]
\caption{Statistics of the two datasets.}
\label{tab:stats}
\centering
\footnotesize
\setlength{\tabcolsep}{4pt}
\begin{tabular}{l l r r r r}
\toprule
\textbf{Dataset} & \textbf{Domain} & \textbf{\#Users} & \textbf{\#Items} & \textbf{\#Inter.} & \textbf{Avg. Len.}\\
\midrule
\multirow{3}{*}{Industrial}
 & Song      & 3.56M & 3.39M & 297.2M & 83.4\\
 & Playlist  & 3.56M & 2.30M & 166.4M & 46.7\\
 & Podcast   & 3.56M & 0.48M & 75.5M  & 21.2\\
\midrule
\multirow{3}{*}{Amazon}
 & Books         & 335.0K & 261.9K & 4.8M & 14.3\\
 & Kindle\_Store & 335.0K & 355.6K & 9.2M & 27.3\\
 & Movies\&TV    & 335.0K &  92.4K & 2.0M &  6.1\\
\bottomrule
\end{tabular}
\end{table}

\subsubsection{Evaluation Protocol}
We focus on the recall task and report HitRate@$K$ (Hit@$K$) and NDCG@$K$ with $K\in\{5,10\}$ on every target domain. On the Industrial Dataset, users are randomly split into training and test sets at a $19{:}1$ ratio; both partitions cover all three domains. For each held-out user, the earliest future interaction in each domain serves as the recall ground truth, and metrics are averaged over users in the test set. The same evaluation protocol is applied to the public benchmark, with a per-domain leave-one-out split: the last interaction in each domain is held out for testing.

\subsubsection{Baselines}
We compare \model against nine representative baselines covering both the single-domain sequential recommendation (SDSR) and MDSR paradigms.

\noindent\textbf{Single-Domain Sequential Recommenders (SDSR).}

GRU4Rec~\cite{hidasi2015session} applies stacked GRUs over the user click sequence for next-item prediction.

SASRec~\cite{kang2018self} uses a unidirectional Transformer encoder under a causal next-item objective.

BERT4Rec~\cite{sun2019bert4rec} pre-trains a bidirectional Transformer with a Cloze-style masked-item objective.

Mamba4Rec~\cite{liu2024mamba4rec} replaces self-attention with a selective Mamba state-space block for linear-time sequential modeling.

\noindent\textbf{Multi-Domain Sequential Recommenders (MDSR).}

MOSAIC~\cite{bahi2026mosaic} disentangles domain-specific, domain-common, and exclusive user components through a triple-encoder with orthogonality and adversarial constraints.

SyNCRec~\cite{park2024pacer} couples single-domain and cross-domain branches via an asymmetric cooperative network reweighted by the per-domain negative-transfer degree.

MDSR-DSL~\cite{hwang2024multi} learns a per-domain representation space and adapts cross-domain item representations through a domain-to-domain mapping module.

AutoCDSR~\cite{ju2025revisiting} formulates cross-domain self-attention as Pareto-optimal multi-objective optimization that suppresses negative transfer.

CGRec~\cite{park2023cracking} uses cooperative-game Shapley values to reweight per-domain losses, complemented by a hierarchical category-to-item contrastive objective.

\subsubsection{Implementation Details}
The hidden dimension is set to $D=64$, and the semantic and ID embeddings use $D_c=D_{\mathrm{id}}=64$. The semantic-ID codebook uses $L=3$ quantization levels with $K_c=256$ codewords per level. Each per-domain Transformer encoder has $2$ layers and $1$ attention head; the shared invariant Transformer has $1$ layer. Per-domain histories are left-padded and truncated to a maximum length of $100$, and the cross-domain mixed history is truncated to $200$. The number of sampled negatives is $K_{\mathrm{neg}}=4{,}095$. The loss weights $\alpha,\beta,\gamma$, and $\mu$ are tuned on the validation set within $[0,1]$. We use Adam with an initial learning rate of $1\!\times\!10^{-3}$, $2{,}000$ warm-up steps, and cosine decay. All experiments run on NVIDIA A100 GPUs.

\subsection{Overall Performance (RQ1)}

Tables~\ref{tab:main-ind} and~\ref{tab:main-pub} report the recall performance of \model and the baselines on the Industrial Dataset and the public Amazon Reviews dataset, respectively. Three observations stand out.

\noindent\textbf{Multi-domain vs.\ single-domain.} Multi-domain methods outperform single-domain ones in general. On the sparse Podcast/Movies\&TV domains, well-designed multi-domain methods (e.g.,\ MOSAIC on Podcast and AutoCDSR on Movies\&TV) achieve the largest gains over the best single-domain model, confirming the value of cross-domain auxiliary signals; this gain, however, does not hold for all multi-domain methods, as shown next.

\noindent\textbf{Limits of existing MDSR.} Under naive multi-domain joint training, negative transfer still arises on the sparse Podcast domain: several MDSR methods (e.g.,\ AutoCDSR and CGRec) even fall below the best single-domain model, indicating that the high-traffic Song domain dominates the shared parameters and that current MDSR methods do not adequately control such cross-domain interference.

\noindent\textbf{Effectiveness of \model.} \model achieves the best results on every metric across both datasets and all three target domains. The improvement is most pronounced on the sparsest domains (Podcast/Movies\&TV), supporting the claim that causal disentanglement protects weaker domains from being dominated.

\begin{table*}[t]
\caption{Overall model performance comparison on the Industrial Dataset. The top two methods are highlighted in bold and underlined.}
\label{tab:main-ind}
\centering
\footnotesize
\setlength{\tabcolsep}{1.5pt}
\tightrules
\begin{tabular}{l*{12}{M}}
\toprule
\multirow{2}{*}{Method} & \multicolumn{4}{c}{Song} & \multicolumn{4}{c}{Playlist} & \multicolumn{4}{c}{Podcast}\\
\cmidrule(lr){2-5}\cmidrule(lr){6-9}\cmidrule(lr){10-13}
& \hd{Hit@5} & \hd{Hit@10} & \hd{NDCG@5} & \hd{NDCG@10} & \hd{Hit@5} & \hd{Hit@10} & \hd{NDCG@5} & \hd{NDCG@10} & \hd{Hit@5} & \hd{Hit@10} & \hd{NDCG@5} & \hd{NDCG@10}\\
\midrule
GRU4Rec   & 0.0253 & 0.0386 & 0.0163 & 0.0205 & 0.0240 & 0.0389 & 0.0150 & 0.0198 & 0.1527 & 0.1977 & 0.1158 & 0.1303\\
SASRec    & 0.0278 & 0.0404 & 0.0193 & 0.0233 & 0.0252 & 0.0414 & 0.0158 & 0.0210 & 0.1624 & 0.2170 & 0.1183 & 0.1360\\
BERT4Rec  & 0.0273 & 0.0401 & 0.0186 & 0.0227 & 0.0201 & 0.0341 & 0.0127 & 0.0172 & 0.1714 & 0.2277 & 0.1230 & 0.1412\\
Mamba4Rec & 0.0310 & 0.0451 & 0.0211 & 0.0256 & 0.0301 & 0.0495 & 0.0190 & 0.0250 & 0.1887 & 0.2454 & 0.1381 & 0.1570\\
\midrule
MOSAIC    & 0.0319 & 0.0461 & 0.0218 & 0.0264 & 0.0313 & 0.0519 & 0.0199 & 0.0265 & \underline{0.2022} & \underline{0.2606} & \underline{0.1519} & \underline{0.1708}\\
SyNCRec   & 0.0340 & 0.0500 & 0.0231 & 0.0283 & \underline{0.0357} & \underline{0.0584} & \underline{0.0226} & \underline{0.0299} & 0.1742 & 0.2327 & 0.1287 & 0.1476\\
MDSR-DSL  & 0.0354 & 0.0504 & \underline{0.0238} & 0.0286 & 0.0337 & 0.0540 & 0.0214 & 0.0279 & 0.1933 & 0.2485 & 0.1438 & 0.1617\\
AutoCDSR  & 0.0328 & 0.0481 & 0.0212 & 0.0261 & 0.0332 & 0.0538 & 0.0209 & 0.0275 & 0.1539 & 0.2060 & 0.1144 & 0.1312\\
CGRec     & \underline{0.0355} & \underline{0.0521} & 0.0237 & \underline{0.0290} & 0.0333 & 0.0545 & 0.0210 & 0.0278 & 0.1536 & 0.2011 & 0.1125 & 0.1278\\
\midrule
\textbf{\model} & \textbf{0.0376} & \textbf{0.0542} & \textbf{0.0261} & \textbf{0.0314} & \textbf{0.0398} & \textbf{0.0636} & \textbf{0.0259} & \textbf{0.0335} & \textbf{0.2622} & \textbf{0.3308} & \textbf{0.1963} & \textbf{0.2185}\\
\midrule
Improv.\,(\%) & +5.9 & +4.0 & +9.7 & +8.3 & +11.5 & +8.9 & +14.6 & +12.0 & +29.7 & +26.9 & +29.2 & +27.9\\
\bottomrule
\end{tabular}
\end{table*}

\begin{table*}[t]
\caption{Overall model performance comparison on the Amazon Reviews Dataset. The top two methods are highlighted in bold and underlined.}
\label{tab:main-pub}
\centering
\footnotesize
\setlength{\tabcolsep}{1.5pt}
\tightrules
\begin{tabular}{l*{12}{M}}
\toprule
\multirow{2}{*}{Method} & \multicolumn{4}{c}{Books} & \multicolumn{4}{c}{Kindle\_Store} & \multicolumn{4}{c}{Movies\&TV}\\
\cmidrule(lr){2-5}\cmidrule(lr){6-9}\cmidrule(lr){10-13}
& \hd{Hit@5} & \hd{Hit@10} & \hd{NDCG@5} & \hd{NDCG@10} & \hd{Hit@5} & \hd{Hit@10} & \hd{NDCG@5} & \hd{NDCG@10} & \hd{Hit@5} & \hd{Hit@10} & \hd{NDCG@5} & \hd{NDCG@10}\\
\midrule
GRU4Rec   & 0.0338 & 0.0464 & 0.0180 & 0.0222 & 0.0171 & 0.0248 & 0.0123 & 0.0150 & 0.0159 & 0.0241 & 0.0109 & 0.0135\\
SASRec    & 0.0379 & 0.0483 & 0.0266 & 0.0299 & 0.0224 & 0.0326 & 0.0165 & 0.0198 & 0.0210 & 0.0315 & 0.0146 & 0.0179\\
BERT4Rec  & 0.0334 & 0.0410 & 0.0213 & 0.0245 & 0.0180 & 0.0261 & 0.0132 & 0.0158 & 0.0168 & 0.0252 & 0.0117 & 0.0143\\
Mamba4Rec & 0.0410 & 0.0517 & 0.0289 & 0.0325 & 0.0255 & 0.0369 & 0.0185 & 0.0221 & 0.0221 & 0.0341 & 0.0156 & 0.0192\\
\midrule
MOSAIC    & 0.0430 & \underline{0.0541} & 0.0298 & \underline{0.0345} & 0.0269 & 0.0383 & 0.0194 & 0.0248 & \underline{0.0224} & \underline{0.0370} & 0.0148 & 0.0190\\
SyNCRec   & \underline{0.0471} & 0.0534 & \underline{0.0317} & 0.0342 & 0.0280 & 0.0397 & 0.0206 & 0.0255 & 0.0209 & 0.0347 & 0.0143 & 0.0185\\
MDSR-DSL  & 0.0402 & 0.0507 & 0.0285 & 0.0332 & 0.0243 & 0.0357 & 0.0176 & 0.0230 & 0.0187 & 0.0294 & 0.0133 & 0.0167\\
AutoCDSR  & 0.0445 & 0.0531 & 0.0312 & 0.0337 & \underline{0.0293} & \underline{0.0411} & \underline{0.0219} & \underline{0.0267} & 0.0210 & 0.0342 & \underline{0.0162} & \underline{0.0207}\\
CGRec     & 0.0391 & 0.0486 & 0.0259 & 0.0298 & 0.0194 & 0.0285 & 0.0136 & 0.0173 & 0.0146 & 0.0221 & 0.0100 & 0.0126\\
\midrule
\textbf{\model} & \textbf{0.0487} & \textbf{0.0555} & \textbf{0.0332} & \textbf{0.0359} & \textbf{0.0310} & \textbf{0.0431} & \textbf{0.0234} & \textbf{0.0283} & \textbf{0.0248} & \textbf{0.0405} & \textbf{0.0181} & \textbf{0.0229}\\
\midrule
Improv.\,(\%) & +3.4 & +2.6 & +4.7 & +4.1 & +5.8 & +4.9 & +6.8 & +6.0 & +10.7 & +9.5 & +11.7 & +10.6\\
\bottomrule
\end{tabular}
\end{table*}

\subsection{Ablation Study (RQ2)}

To quantify the contribution of every component, the following variants are constructed.
\begin{itemize}
\item \textbf{w/o Ortho}: removes the orthogonality regularizer $\mathcal{L}_{\mathrm{ortho}}$.
\item \textbf{w/o Adv}: removes the GRL-based domain adversarial loss $\mathcal{L}_{\mathrm{adv}}$.
\item \textbf{w/o Causal}: removes both orthogonality and adversarial losses, reducing \model to a multi-task baseline without causal disentanglement.
\item \textbf{w/o Multi-modal}: removes all multi-modal information (text, vision, and audio) and the semantic-ID codebook, relying only on domain-specific ID embeddings.
\end{itemize}

\begin{table*}[t]
\caption{Ablation results on the Industrial Dataset.}
\label{tab:ablation}
\centering
\footnotesize
\setlength{\tabcolsep}{1.5pt}
\tightrules
\begin{tabular}{l*{12}{M}}
\toprule
\multirow{2}{*}{Variant} & \multicolumn{4}{c}{Song} & \multicolumn{4}{c}{Playlist} & \multicolumn{4}{c}{Podcast}\\
\cmidrule(lr){2-5}\cmidrule(lr){6-9}\cmidrule(lr){10-13}
& \hd{Hit@5} & \hd{Hit@10} & \hd{NDCG@5} & \hd{NDCG@10} & \hd{Hit@5} & \hd{Hit@10} & \hd{NDCG@5} & \hd{NDCG@10} & \hd{Hit@5} & \hd{Hit@10} & \hd{NDCG@5} & \hd{NDCG@10}\\
\midrule
\textbf{Full \model} & \textbf{0.0376} & \textbf{0.0542} & \textbf{0.0261} & \textbf{0.0314} & \textbf{0.0398} & \textbf{0.0636} & \textbf{0.0259} & \textbf{0.0335} & \textbf{0.2622} & \textbf{0.3308} & \textbf{0.1963} & \textbf{0.2185}\\
w/o Ortho        & 0.0359 & 0.0518 & 0.0251 & 0.0303 & 0.0356 & 0.0594 & 0.0232 & 0.0309 & 0.2386 & 0.3010 & 0.1786 & 0.1988\\
w/o Adv          & 0.0356 & 0.0515 & 0.0248 & 0.0299 & 0.0371 & 0.0610 & 0.0241 & 0.0319 & 0.2438 & 0.3076 & 0.1826 & 0.2032\\
w/o Causal       & 0.0351 & 0.0508 & 0.0243 & 0.0292 & 0.0349 & 0.0585 & 0.0227 & 0.0303 & 0.2281 & 0.2878 & 0.1708 & 0.1901\\
w/o Multi-modal  & 0.0331 & 0.0477 & 0.0227 & 0.0273 & 0.0318 & 0.0508 & 0.0206 & 0.0266 & 0.2063 & 0.2601 & 0.1523 & 0.1695\\
\bottomrule
\end{tabular}
\end{table*}

\noindent Table~\ref{tab:ablation} shows that: (i) removing either the orthogonality regularizer (w/o Ortho) or the adversarial loss (w/o Adv) alone consistently degrades performance across all three domains, showing that each causal constraint individually contributes to representation disentanglement; (ii) removing both causal constraints (w/o Causal) yields a further drop, confirming that orthogonality and domain adversarial training are complementary; (iii) removing the multi-modal stream causes the largest overall degradation across all three domains, with Podcast suffering the most, confirming that visual and audio semantics are critical for bridging content-heterogeneous domains.

\subsection{Cross-Domain Cold-Start Analysis (RQ3)}

\model is able to perform cold-start prediction for users who are cold in a target domain by transferring knowledge from their behaviors in other domains. To this end, we design a strict cross-domain cold-start protocol on the Industrial Dataset. We randomly split the test set into three equal folds, each assigned a single cold-start target domain. For every user, we mask \emph{all} interactions belonging to their assigned target domain from the cross-domain history and keep only the behaviors from the other two domains; the user thus has zero history in the target domain and is equivalent to a genuine cold-start user there. The masked target-domain ground-truth item is then ranked over the full item vocabulary of that domain. All compared models use the same user-fold split for a strictly fair comparison, and performance is measured by Hit@$K$ and NDCG@$K$ with $K\in\{5,10\}$.

\begin{figure*}[t]
  \centering
  \includegraphics[width=\textwidth]{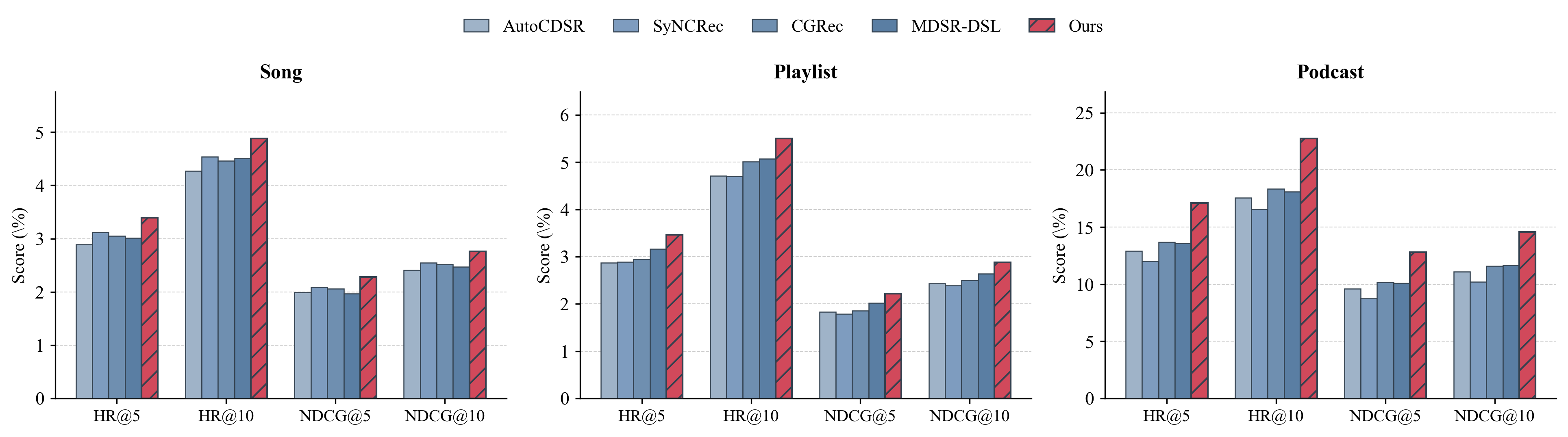}
  \caption{Cross-domain cold-start performance (\%) across three target domains. Users are cold in the target domain and ranked over its full item vocabulary using only their behaviors from the other two domains.}
  \label{fig:coldstart}
\end{figure*}

Fig.~\ref{fig:coldstart} reports the per-domain results. \model consistently outperforms the strongest MDSR baselines on all three target domains. Because the target-domain history is entirely absent, the gains come purely from cross-domain knowledge transfer: TransNet completes the missing target-domain representation at the $\tilde{\mathbf{h}}$ level, while the shared invariant Transformer and the domain adversarial loss keep the domain-invariant preference domain-agnostic and therefore transferable to the unseen target domain. The advantage is most pronounced on the traffic-sparse Podcast domain, confirming the effectiveness of the proposed approach under user cold-start conditions.

\subsection{Online A/B Test (RQ4)}

In May 2026, we conducted a two-week online A/B test on NetEase Cloud Music, a large-scale audio streaming platform delivering song, playlist, and podcast recommendations to tens of millions of users daily. The control group adopted the existing online per-domain recall pipeline as the baseline; the treatment group enhanced the baseline by additionally introducing \model as a new recall channel, while sharing identical candidate-ranking and re-ranking modules with the control group. Our core metrics are user engagement behaviors in the main traffic scenario of each domain, specifically \emph{play duration} on Song and Podcast and \emph{like rate} on Playlist. The \emph{play duration} metric measures the average listening time users spend per session, while the \emph{like rate} metric measures the proportion of exposed playlists that users add to their favorites. During the trial period, the improvements of the treatment group over the control group across all three domains are summarized in Table~\ref{tab:online}, and all reported gains are statistically significant. The gains are most pronounced on the traffic-sparse Podcast domain, consistent with the offline findings and confirming the practical value of \model at industrial scale.

\begin{table}[!ht]
\caption{Online A/B test improvements over the control group.}
\label{tab:online}
\centering
\footnotesize
\setlength{\tabcolsep}{6pt}
\begin{tabular}{l l c}
\toprule
\textbf{Domain} & \textbf{Metric} & \textbf{Improvement}\\
\midrule
Song     & Play Duration & $+1.0\%$\\
Playlist & Like Rate     & $+3.0\%$\\
Podcast  & Play Duration & $+1.6\%$\\
\bottomrule
\end{tabular}
\end{table}

\section{Conclusion}\label{sec:conclusion}

This paper introduces \model, a multi-modal causal disentanglement framework for multi-domain sequential audio recommendation. By revisiting multi-domain preference generation from a structural causal perspective, \model decomposes user representations into a domain-invariant preference and per-domain specific preferences, and enforces three complementary constraints---orthogonality between the two factors, domain adversarial training on the invariant factor, and architectural restriction of cross-domain interaction---to suppress negative transfer in joint optimization. A trainable semantic-ID codebook integrates pre-trained text, vision, and audio features into a shared embedding space, and a transfer network mitigates cold start through pre-disentanglement imputation. Offline experiments on a production log from NetEase Cloud Music and a public benchmark, together with a two-week online A/B test, validate that \model improves recommendation performance and is particularly effective on traffic-sparse domains.

\section*{Acknowledgment}

This research was supported by the Natural Science Foundation of Zhejiang Province under Grant No.~LZ25F020010 and the National Natural Science Foundation of China under Grant No.62672160.

\bibliographystyle{IEEEtran}
% Balance the final page while keeping reference 27 in one column.
\IEEEtriggeratref{27}
\bibliography{references}

\end{document}